\documentclass[nofootinbib,aps]{revtex4}

\usepackage{amssymb,amstext,amsmath,amsthm}
\usepackage[dvips]{graphicx}
\usepackage{latexsym}
\usepackage{psfrag}
\usepackage{amsfonts}
\usepackage{bbm}
\usepackage{axodraw}
\usepackage{color}

\newcommand{\Ref}[1]{(\ref{#1})}

\newcommand{\eqa}{\begin{eqnarray}}
\newcommand{\neqa}{\end{eqnarray}}
\newcommand{\equ}{\begin{equation}}
\newcommand{\nequ}{\end{equation}}
\newcommand{\no}{\nonumber\\}
\newcommand{\be}{\begin{equation}}
\newcommand{\ee}{\end{equation}}

\newtheorem{theo}{Theorem}

\newcommand{\R}{\mathbb{R}}

\def\la{\langle}
\def\ra{\rangle}
\newcommand{\bra}[1]{\la {#1}|}
\newcommand{\ket}[1]{|{#1}\ra}
\newcommand{\mean}[1]{\la{#1}\ra}

\newcommand{\1}{$\{10j\}$}

\def\d{\delta}
\def\f{\frac}
\def\tl{\tilde}

\newcommand{\p}{\partial}
\newcommand{\n}{\nabla}

\usepackage{bbm}
\def\C{{\mathbbm C}}

\newcommand{\SU}{\mathrm{SU}}
\newcommand{\SO}{\mathrm{SO}}
\newcommand{\SL}{\mathrm{SL}}

\let\eps=\epsilon

\newcommand{\lp}{\ell_{\rm P}}

\newcommand{\Spin}{\mathrm{Spin}}
\def\Ee{{\cal E}}
\def\kk{{\cal K}}

\def\vphi{\varphi}
\def\tl{\tilde}

\begin{document}

\title{{\large\bf Group Integral Techniques for the Spinfoam Graviton Propagator}}
\author{{\bf Etera R. Livine}\footnote{elivine@perimeterinstitute.ca} }
\affiliation{Laboratoire de Physique, ENS Lyon, CNRS UMR 5672, 46 All\'ee d'Italie, 69364 Lyon,
France, and \\
Perimeter Institute, 31 Caroline St. N, Waterloo, ON N2L 2Y5, Canada.}
\author{{\bf Simone Speziale}\footnote{sspeziale@perimeterinstitute.ca}}
\affiliation{Perimeter Institute, 31 Caroline St. N, Waterloo, ON N2L 2Y5,
Canada.}

\begin{abstract}{
\noindent {We consider the recent proposal \cite{RovelliProp} for the extraction of the graviton
propagator from the spinfoam formalism. We propose a new ansatz for the boundary state,
using which we can write the propagator as an integral over $\SU(2)$. The
perturbative expansion in the Planck length can be recast into the saddle point expansion
of this integral. We compute the leading order and recover the behavior expected from
low--energy physics. In particular, we prove that the degenerate spinfoam configurations
are suppressed.}}
\end{abstract}

\maketitle



\section{Introduction}

Recently \cite{ModestoProp, RovelliProp, Io, noi}, a proposal has appeared for the
computation of the 2-point function of quantum gravity within the spinfoam
formalism \cite{carlo}, a candidate covariant approach to a
non--perturbative quantisation of General Relativity (GR).
This proposal offers a possibility to study the semiclassical limit of spinfoams
and define the perturbative expansion in the Planck length $\lp$, arguably the major open
question within this approach.
This is very interesting, for two different reasons: on the one hand,
to check the correctness of the low--energy limit of
spinfoams; on the other hand, to address the possibility of curing the non--renormalisability of the conventional
perturbative expansion via background--independent methods.

In \cite{RovelliProp} (see also \cite{noi}), the proposal
was applied to the Barrett--Crane (BC) spinfoam model for four dimensional (4d)
Riemannian GR, and it was shown that the 2-point function, or graviton propagator, indeed had the correct
$1/|x-y|^2$ leading order behavior in the large scale limit. However, there are some
assumptions behind this result that deserve a more careful treatment. In particular, the
BC model is plagued by degenerate configurations. This has so far cast some doubts on the
viability of the proposal, especially because there is still no support from the numerical analysis,
due to the high complexity of the vertex amplitude.

In this work, we prove that the degenerate configurations do not affect the leading order.
In doing so, we modify the ansatz for the boundary state, following
the 3d investigation appeared in \cite{noi2}. For
the 4-simplex spinfoam contribution, the new boundary state allows us to
write the 2-point function as an integral over $\SU(2)$. This is our first result.
In the large $j$ limit, we evaluate
this integral in a saddle point approximation, proving the $1/|x-y|^2$ behavior of the
leading order. The degenerate configurations can be correctly neglected because they correspond to
saddle points which are not absolute minima.
This is our second result. Thanks to the integral expression, which involves no sums,
the numerical analysis is  strongly simplified. Indeed, the leading order result is fully
supported by  numerical analysis \cite{Dan}.

Concerning the full perturbative expansion in $\lp$, this can be computed from higher
orders of the saddle point approximation as well as from the contribution of other
spinfoams. We do not attempt here their evaluation, which is rather complicate. Notice
however that higher orders have been studied in a 3d toy model \cite{Io, noi2};
remarkably, interesting modifications to the conventional expansion arise, due to the
microscopical quantum geometry described by spinfoams. This makes the whole approach very
interesting, and pushes towards the calculation of these corrections in the 4d case. We
expect that having recast the $\lp$ expansion into the saddle point expansion will help
future work.

This paper is organised as follows. In the next Section, we briefly recall properties of the BC vertex
amplitude which we need in the rest of the paper. In Section \ref{Section2}, we review the construction
of the 2-point function and the results of \cite{RovelliProp}. In Section \ref{SectionBoundary}, we introduce the
new boundary state and discuss its properties. In Section \ref{SectionProp}, we show how the new boundary state allows
to write the 2-point function as an integral over $\SU(2)$. In Section \ref{SectionSaddle}, we study the saddle
point approximation to this integral, and show that the leading order behaves as expected from low--energy physics.
In the final Section, we discuss possible further developments.

Throughout the paper, we use units $\lp=1$.

\section{Spinfoam amplitudes: evaluation of Relativistic Spin Networks}

In this Section we recall basic facts about the spinfoam amplitudes, the quantities
encoding the dynamics of quantum gravity. This gives us the opportunity to fix our
notation and to pinpoint features which will be crucial in the following. We consider
here the spinfoam amplitude for the BC model of 4d Riemannian quantum gravity; this is
given by the evaluation of a relativistic spin network  \cite{BC} with group $\Spin(4)$,
namely the double cover of $\SO(4)$. The relativistic spin network is defined by a graph
$\Gamma$ together with the assignment of a $\Spin(4)$ group element $G_n$ to each node
$n$ of $\Gamma$ and a simple irreducible representation (irrep) $J_l$ to each link $l$ of
$\Gamma$. Using the homomorphism $\Spin(4)=\SU(2)_L\times\SU(2)_R$, the irreps of
$\Spin(4)$ are labelled by two half--integers, say $(j, k)$, corresponding to the irreps
of the two $\SU(2)$ sectors; then the simple representations are such that they induce
the same $\SU(2)$ representation in the left and right sectors, namely $J_l=(j_l,j_l)$.
Furthermore, let us note here that the scalar Casimir of $\Spin(4)$ satisfies
$C^2_{\Spin(4)}(j,k)=4\, [C^2(j)+C^2(k)]$, where $C^2(j)$ is the $\SU(2)$ Casimir.

Using the above homomorphism, each group element decomposes as the product of two left and
right rotations $G=g_Lg_R$, and the evaluation reads \cite{Barrett},
\be
\Ee_{\Gamma}\,\equiv\,
\int_{\Spin(4)} \prod_n dG_n\,
\prod_l \kk_{J_l}(G_{s(l)}^{-1}G_{t(l)}),
\ee
where $s(l)$ and $t(l)$ denote the source and target node of the link $l$. The kernel
$\kk_J(G)$ is the matrix element of $G$ on the $\SU(2)$-invariant vector $|J,0\ra$ in the
$J$ representation. Here $\SU(2)$ is the diagonal rotation group, corresponding to the
subgroup of 3d rotations.
We conveniently parametrise $\SU(2)$ group elements as
\equ\label{g}
g(\phi, \hat n) = \cos\phi \, \mathbbm{1}+ i \, \sin\phi \, \hat n\cdot \vec\sigma, \quad \phi\in[0, \pi].
\nequ
Consequently, the characters are given by $\chi_j(g)=\f{\sin d_j\phi}{\sin\phi}$,
and the Haar measure is $dg = \f1{2\pi^2} \sin^2\phi\, d^2\Omega(\hat n)\, d\phi$.

The invariant vector is easily expressed in term of left/right
components:
\be
|J,0\ra\,=\,
\f{1}{\sqrt{d_j}}\sum_m (-1)^{j-m}\,|j, m\ra_L\,|j, -m\ra_R
\,=\,
\f{1}{\sqrt{d_j}}\sum_m |j, m\ra_L\,{}_R\la j, m|,
\ee
where $d_j=2j+1$ is the dimension of the $\SU(2)$ representation of spin $j$. Then it
is straightforward to realize that the $\SU(2)$ invariant kernel $\kk_J$ is simply the
$\SU(2)$ character:
\be
\la J,0|G|J,0\ra
\,=\,
\la J,0|g_Lg_R|J,0\ra
\,=\,
\f{1}{d_j}\sum_m \la j, m|g_Lg_R^{-1}|j, m\ra
\,=\,
\f1{d_j}\chi_j(g_Lg_R^{-1}).
\ee
Finally, using the invariance of the Haar measure $dG=dg_L dg_R$ under left and right
multiplication, it is easy to prove that the relativistic spin network evaluation is
actually a 3d object regarding only integrals over $\SU(2)$:
\be\label{eps}
\Ee_\Gamma\,=\,
\int_{\SU(2)} \prod_n dg_n\,
\prod_l \f1{d_{j_l}}\chi_{j_l}(g_{s(l)}^{-1}g_{t(l)}),
\ee
In particular, the vertex amplitude for the BC model is obtained for $\Gamma$ given by a
4-simplex, and this gives the \1 symbol for the recoupling theory of $\SU(2)$,
\equ\label{10jint}
\{10j\}= \int_{\SU(2)} \prod_n dg_n \, \prod_l \f1{d_{j_l}}{\chi}_{j_l}(g_{s(l)}^{-1}g_{t(l)}).
\nequ
Let us point out that the \1 symbol is defined up to a normalisation. This creates
an ambiguity in the definition of the BC model.
The standard normalisation found in the literature is  \Ref{10jint}
without the $1/d_j$ factors. Notice also that another natural choice
for the kernel is $d_j^2\kk_J=d_j\chi_j$, instead of $\chi_j/d_j$, as
it normalises the convolution product:
$$
\int_{\Spin(4)}dG\, \left(d_j^2\kk_J(HG^{-1})\right)\,
\left(d_j^2\kk_J(GK^{-1})\right)\,=\,
d_j^2\kk_J(HK^{-1}).
$$

These normalisation issues do not modify the computations below, namely the leading order of the
graviton correlation computed using a single 4-simplex. We nevertheless mention them as they will most
likely affect the higher order corrections and become essential
when considering configurations with many 4-simplices.

\subsection*{Geometrical interpretation of the \1 symbol}

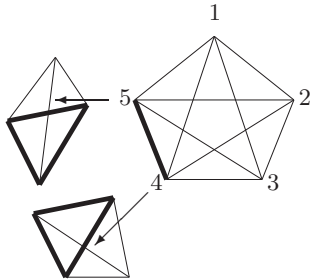
\begin{figure}
\begin{picture}(0,0) (0,0)
\SetWidth{0.5}\SetScale{0.6}\SetColor{Black}
\Line(-50,0)(50,0)\Line(-50,0)(0,40)\Line(50,0)(0,40)\Line(-30,-50)(0,40)\Line(30,-50)(0,40)
{\SetWidth{3}\Line(-30,-50)(-50,0)}\Line(30,-50)(50,0)\Line(-30,-50)(50,0)\Line(30,-50)(-50,0)
\Line(-30,-50)(30,-50)\put(-2,30){1}\put(32,-2){2}\put(-36,-2){5}\put(20,-34){3}\put(-24,-34){4}
\put(-40,0){\vector(-1,0){20}}\put(-25,-35){\vector(-1,-1){20}}
\end{picture}
\begin{picture}(0,0)(-13,25)\SetWidth{0.5}\SetScale{0.6}\SetColor{Black}
{\SetWidth{3}\Line(-140,-30)(-90,-20)\Line(-120,-70)(-90,-20)\Line(-120,-70)(-140,-30)}
\Line(-140,-30)(-80,-70)\Line(-90,-20)(-80,-70)\Line(-120,-70)(-80,-70)
\end{picture}
\begin{picture}(0,0)(0,-10)\SetWidth{0.5}\SetScale{0.6}\SetColor{Black}
{\SetWidth{3}\Line(-140,-30)(-90,-20)\Line(-120,-70)(-90,-20)\Line(-120,-70)(-140,-30)}
\Line(-140,-30)(-110,10)\Line(-90,-20)(-110,10)\Line(-120,-70)(-110,10)
\end{picture}
\vspace{2cm}
\caption{The 4-simplex (or pentahedral) boundary spin network. We label the nodes $I=1,\ldots 5$. In the dual picture,
they are in correspondence with tetrahedra of the boundary triangulation. Two of
them are represented. The links $IJ$,
on the other hand, are dual to triangles. Consider for instance the link $45$: this is dual
to the triangle shared by the tetrahedra 4 and 5. Associated with the link 45 is the dihedral
angle $\phi_{45}$ between the tetrahedra 4 and 5. \label{4simplex}}
\end{figure}

The \1 symbol admits a geometrical interpretation, associated to the structure of a 4-simplex,
which will be important in the following. The key fact is that
it can be written as an integral over ten $\SU(2)$ angles $\phi_l\in[0, \pi]$,
\be\label{10j}
\{10j\}= \int d\mu[\phi_l] \, \prod_l \f1{d_{j_l}}{\chi}_{j_l}(\phi_l),
\ee
where the measure takes into account that the angles come from the
vertex group elements $g_v$ through the relation, $\cos{\phi_l}= \f12 {\rm
tr}(g_{s(l)}^{-1}g_{t(l)})$.
Let us introduce the notation $(IJ)$ for $l$ linking the
nodes $I$ and $J$, such that $\phi_{IJ}\equiv \phi_l$, with the convention $\phi_{II}=0$.
The above relation imposes a constraint that can be written as the vanishing of the determinant
of the 5$\times$5 Gram matrix $G_{IJ}=\cos\phi_{IJ}$, as shown in
\cite{asymptlaurent}:
\be\label{ten}
d\mu[\phi_l]\,=\,
\prod_l d\phi_l \,\sin{\phi_l}\,\delta\Big(\det G_{IJ}\Big).
\ee
This constraint has a clear geometrical interpretation: it
says that the angles $\phi_l$ are the dihedral
angles of a certain 4-simplex. Indeed, notice that the spin network induces a dual triangulation
which is also a 4-simplex, with tetrahedra dual to the nodes and triangles dual to the links (see Fig.\ref{4simplex}).
Then the constraint can be translated into the Schl\"afli identity
$\sum_{l} A_{l}(\phi)\,d\phi_l \,=\,0,$
where $A_l$ is the area of the triangle (dual to the link $l$) of the geometric 4-simplex,
and $\phi_l$ its dihedral angle.
The areas can be written as derivatives of the Gram matrix,
$A_l = \kappa \, \f{\p {\det G_{IJ}}}{\p \phi_l},$ where $\kappa$ is a  proportionality constant, related
to the 4-volume of the simplex.
For more details see the Appendix.

The original group integration in \Ref{10jint} is
over 5 copies of the 3-sphere ${\cal S}^3\sim\Spin(4)/\SU(2)$ and the 10 angles
$\phi_l$ are easily related to the 5 original 4d unit vectors $\hat{N}_I\in{\cal S}^3$, via
$\cos{\phi_{IJ}}=\hat{N}_I \cdot \hat{N}_J$.

For later use, let us consider the equilateral case when all the dihedral angles are equal, $\phi_l=\vphi$. Then the
constraint simply reads
\be
\det G_{IJ}(\phi_l=\vphi)=(1-\cos\vphi)^4\,(1+4\cos\vphi),  
\ee
whose roots are given by 
the completely degenerate 4-simplex $\vphi=0$, and the equilateral 4-simplex $\vphi\equiv\theta=\arccos(-1/4)$.

To summarise, two important features emerge from the above discussion:
\begin{itemize}
\item{The relevant group for 4d Riemannian quantum gravity without matter is $\SU(2)$.}
\item{The angles $\phi$ parametrising $\SU(2)$ group elements can be thought of as dihedral
angles between the tetrahedra dual to the nodes of the boundary spin network entering the
spinfoam amplitude.}\end{itemize}
Equipped with these considerations, we now proceed to describe how to construct the 2-point function.

\section{The 2-point function}
\label{Section2}
In the conventional quantum field theory framework
one expands the metric tensor around a given background, typically the Minkowski flat
spacetime $\eta_{\mu\nu}$, writing $g_{\mu\nu} = \eta_{\mu\nu}+ \lp h_{\mu\nu}$;
then the 2-point function,
\equ\label{W}
W_{\mu\nu\rho\sigma}(x,y)=\bra{0} \,{\rm T}
\left\{h_{\mu\nu}(x) h_{\rho\sigma}(y)\right\} \ket{0},
\nequ
is evaluated in perturbation theory. The leading order, coming from the quadratic term in
the action, goes as  $1/|x-y|^2$, namely as $1/p^2$ in momentum space. The higher orders
in the action give self--energy corrections.

In the spinfoam formalism, the 2-point function \Ref{W} can be studied looking at the correlations
between fluctuations of geometrical quantities. The fluctuations are defined with respect
to a boundary metric which is encoded in a spin network state $s$. The points $x$ and $y$ in
\Ref{W} can be identified with nodes of $s$, and the directions $\mu$ can be defined
using the links of $s$. In particular, since each link of $s$ can be thought of as
dual to a triangle in a triangulation of the boundary, it is convenient to project
\Ref{W} along the normals to these triangles. Namely,  using the boundary metric,
we contract the indices of \Ref{W} with four normal vectors, two belonging to the
(3d region dual to) the node $x$ and two to the node $y$.
For each node, we should distinguish the case when we take twice the same normal,
from the case when we consider two different normals. In the first case, the contraction
with $h_{\mu\nu}$ gives (the fluctuation of) the area of the triangle. In the second case, of
the dihedral angle between the two triangles.

In the quantum theory, both areas and dihedral angles are represented by quantum
operators. The area operator has spectrum given by twice the square root of the $\SU(2)$
Casimir operator,\footnote{The factor 2 between the triangle area and the Casimir can be
understood as follows. Given the two edge vectors $a^I$, $b^J$ of the triangle, the
$\Spin(4)$ Casimir can be identified with the norm of the bivector $B^{IJ} = \f12(a^I b^J
- b^I a^J)$, namely with the quantity $\f12 |a|^2 |b|^2 \sin^2\theta \equiv 2 A^2$.
Therefore $2 A^2=C^2_{\Spin(4)}(j,j) \equiv 8 \, C^2(j)$, using the homomorphism between
$\Spin(4)$ and $\SU(2)$ introduced above. As for the spectrum, let us recall that the
Casimir operator is always defined up to an additive constant. Usually one takes this
constant to zero, so that its spectrum reads $C^2(j)=j(j+1)$. On the other hand, here we
take a shift of $1/4$, so that $C^2(j)=(j+\f12)^2$, to match the results on the
asymptotics of the
\1 symbol reported below.} $A=2 \, C(j)=2j+1$.
Analogously, the angle operator can be expressed in terms of $\SU(2)$ Casimir operators,
via the recoupling theorem (see for instance \cite{noi, semi}). This is how the spin
variables entering the amplitudes are related to geometrical quantities. Since the
fundamental variables of the theory are triangle areas and dihedral angles, we compute
the 2-point function as the correlation between fluctuations of these quantities around a
flat background. Choosing a flat background for the boundary geometry allows us to
introduce a (spin) scale parameter, which we call $j_0$. We do so by describing the
triangulation of the flat boundary in terms of equilateral triangles with areas $A_0 = 2
\, C(j_0)$. The parameter $j_0$ can be used to measure the physical distance between two
points on the boundary with respect to the flat background metric.

In the following, we consider only the correlation between the areas, leaving a discussion
of the other cases for the conclusive Section.
Introducing the normal vectors $n^\mu_a$ and $n^\mu_a$ to the triangles $a$
and $b$, the area correlations can be compared with the
projections $n^\mu_a n^\nu_a n^\rho_{b} n^\sigma_{b} \, W_{\mu\nu\rho\sigma}(x, y)$ of the
continuum 2-point function.
Notice that if the 4-simplex is equilateral, there are only three independent
projections \cite{noi};
these correspond to the three cases when the triangles $a$ and $b$ are the same, when they share a side,
or when the share only a point.
Following \cite{RovelliProp}, the area correlations are given by \equ\label{W2}
W_{ab}(j_0)=\f1{j_0^4}\f1{\cal N} \ \sum_{s}  \Psi_0[s]
\ {\mathbbm h}(j_a) \ {\mathbbm h}(j_b) \ {\cal A}[s],
\nequ
where ${\cal N}=\sum_{s} \Psi_0[s] \ {\cal A}[s]$ is the normalisation,
and the factor $1/j_0^4$ comes from normalising the projections along the normals.

Let us briefly explain this formula, referring to \cite{noi} for a more complete
description. The graviton propagator is expressed as a sum over all possible boundary
spin networks. This includes a sum over all the the graphs and a sum over all possible
assignments of spins to the links of the graphs.

The boundary state $\Psi_0[s]$ should represent a coherent semiclassical state
of the boundary geometry \cite{noi}. In particular,
we require that the relative uncertainties of the
geometry on this state vanish in the large $j_0$ limit, namely
\equ\label{semi}
\f{\bra{\Psi_0}{\Delta j_l}\ket{\Psi_0}}{\bra{\Psi_0}{j_l}\ket{\Psi_0}} \mapsto 0, \qquad
\f{\bra{\Psi_0}{\Delta \phi_l}\ket{\Psi_0}}{\bra{\Psi_0}{\phi_l}\ket{\Psi_0}} \mapsto 0
\nequ
for all links $l$ in the boundary spin network.

The quantity $\mathbbm h(j_a) := C^2(j_a)-C^2(j_0)$ represents
the field insertion $h_{\mu\nu}(x)$ in a Coulomb--like gauge--fixing \cite{Io}. We see that it
is (one fourth) the fluctuation of the squared area.

The amplitude ${\cal A}[s]$ is given by
the spinfoam model chosen, and it is a sum over all spinfoams $\sigma$ whose boundary $\p
\sigma$ is given by the spin network $s$. Working with the Barrett-Crane model, we have
\equ\label{bc}
{\cal A}[s] = \sum_{\sigma \atop \p \sigma = s}
\sum_{j_f} \; \prod_f d_{j_f}^2 \,
\prod_{e\in\sigma\setminus\partial\sigma} A_e(j_f)
\prod_{e\in\partial\sigma} (A_e(j_f))^{\f12}
\,\prod_v \, \{10j\},
\nequ
where $v,f$ are respectively the vertices and faces of the spinfoam,  and $j_f\equiv j_l$
whenever a face $f$ intersects the boundary forming a link $l$. The vertex amplitude is
given by the \1 symbol introduced in the previous Section.
Notice that we left unspecified the edge amplitude $A_e(j_f)$.
This reflects the normalisation ambiguity of the \1 symbol,
and different choices lead to different versions of the BC model\footnotemark.
However we expect changes in the edge amplitudes not to affect the leading order of the graviton correlations
$W_{ab}$, in analogy with the 3d case \cite{noi2}.
They will nevertheless affect the higher order corrections, allowing
to distinguish and discriminate the different choices of $A_e$. Below we
will choose a particular edge amplitude which simplifies the calculations of the
leading order.

\footnotetext{ \label{d4} In the literature,
the edge (or tetrahedron) amplitude $A_e(j_1,j_2,j_3,j_4)$ is usually  the product of
some particular powers of $d_{j_1}d_{j_2}d_{j_3}d_{j_4}$ and
$d_{j_1j_2j_3j_4}$, where the latter
is the dimension of the intertwiner space and is given by the norm of
the Barrett-Crane intertwiner:
$$
d_{j_1j_2j_3j_4}=\int_{\SU(2)} dg \,\prod_{i=1}^4\chi_{j_i}(g).
$$ }

The expression \Ref{W2} is the full graviton propagator. This can be evaluated in
perturbation theory. The parameter of the conventional perturbative expansion is the
Planck length $\lp$. By dimensional analysis, it appears (squared) in front of $j_0$. We
are then led to consider the limit $\lp\mapsto 0$, $j\mapsto\infty$ such that $\lp^2 j$
is constant, to study the semiclassical behavior of the 2-point function. This idea is
supported by the asymptotic behavior of the \1 symbol for large spins, which we recall
here.

In the homogeneous large spin limit, namely when
$j_l=N k_l$ and $N\mapsto\infty$, the integral \Ref{10jint} defining the \1 symbol can be
evaluated with a saddle point approximation, leading to the
asymptotics \cite{asymptjohn, asymptlaurent}
\equ\label{asymp}
\{10j\}\sim \sum_\tau \, P(\tau) \, \cos\left(S_{\rm R}[j_l]+\kappa_\tau \f\pi4\right)+D(j_l).
\nequ
Here $S_{\rm R}[j_l]=\sum_{l}d_{j_l} \theta_l$ is the Regge action associated with the
4-simplex, dual to the foam vertex, with triangle areas $A_l=d_{j_l}$; $P(\tau)$ and
$\kappa_\tau$ are factors depending on the combinatorial structure of the 4-simplex, and
$D(j_l)$ is a contribution coming from degenerate configurations of the 4-simplex. In
principle, the emergence of the Regge action supports the idea that the large $j$ limit
can be used to study the semiclassical properties of the theory. However, the degenerate
term $D(j_l)$ dominates strongly the asymptotics, thus hiding the physically interesting
first term of \Ref{asymp}. Indeed, the numerical analysis
in \cite{asymptbaez} could not confirm \Ref{asymp}.
This fact has raised doubts over the BC model. Nevertheless,
it was suggested in \cite{RovelliProp} that the sick term $D(j_l)$ is in fact negligible
in the computation of \Ref{W2}, because it does not match the boundary data induced by
$\Psi_0[j_l]$. Before showing how this happens, let us discuss the structure of the
perturbative expansion.

Using the approximation \Ref{asymp} in \Ref{W2}, we compute the 2-point function within a
particular Regge path integral formulation of quantum gravity. The measure for this path
integral is induced by the spinfoam formalism. On the other hand, the Regge action is a
discretised version of GR, and it can be expanded around the flat background, as in the
continuum. The quadratic term in the action is responsible for the behavior of the
leading order of the 2-point function. Higher order terms in the action give corrections
in powers of $\lp$. However, because \Ref{asymp} is only an approximation, we expect to
have higher order corrections to the 2-point function which are not described by Regge
calculus. Indeed, this expectation is confirmed in 3d \cite{noi2}, where an interesting
structure of the corrections emerges.

This is not the end of the story. For any $s$ in \Ref{W2}, the amplitude
\Ref{bc} sums over all spin foams interpolating $s$. Using as in \cite{RovelliProp}
the group field theory generated
BC model, each foam is weighted by a factor $\lambda^V$, where $V$ is the total number
of vertices in $\sigma$, and $\lambda$ is the dimensionless coupling constant of
the group field theory. Assuming that the perturbative expansion in $\lambda$ is well defined,
this suggests that from the group field theory point of view,
the dominant contributions come from the simplest foams.
Indeed, at first order in $\lambda$ we have a single contribution to \Ref{W2} (see \cite{noi}),
whose boundary is the 4-simplex spin network.

Understanding the precise interplay between the $\lp$ expansion and the $\lambda$
expansion is beyond the scope of this paper. In the rest of this paper, we focus on this
single contribution. See the conclusions for more comments.

\subsection*{The Gaussian boundary state and the leading order of the perturbative expansion}

Let us recall here the contribution to the 2-point function coming from a
single 4-simplex, as computed in \cite{RovelliProp}.
When we restrict the boundary $s$ to be only the pentahedral spin network,
\Ref{W2} reduces to
\equ\label{W2bis}
W_{ab}(j_0)=\f1{j_0^4}\f1{\cal N} \ \sum_{j_l} {\cal A}[j_l] \ \Psi_0[j_l]
\ {\mathbbm h}(j_a) \ {\mathbbm h}(j_b).
\nequ
To explicitly define ${\cal A}[j_l]$, we need to make a choice for the edge amplitudes
$A_e$. The amplitude ${\cal A}[j_l]$ will then involve the face factors $d_j^2$ and the
square-root of the boundary edge amplitudes $(A_e)^{1/2}$. We choose $A_e(j_f)=(\prod_{f}
d_{j_f})^{-1}$.
Taking into account the five tetrahedra of the boundary of the 4-simplex,
we have $\prod_{e\in\p\sigma} \sqrt{A_e(j_f)}=\prod_{f} \prod_e
d_{j_f}^{-1/2}\equiv \prod_f d_{j_f}^{-1}$ and thus:
\equ\label{cicciopanza}
{\cal A}[j_l] \,=\, \prod_l  d_{j_l}^2 \, \prod_l \f1{d_{j_l}}\,\{10j\}
\,=\, \prod_l d_{j_l}\{10j\}
\,=\, \int \prod_n dg_n \, \prod_l \chi_{j_l}(g_{s(l)}^{-1}g_{t(l)}).
\nequ

To compute the free propagator, namely the leading order in the large $j_0$ expansion of
\Ref{W2}, it is sufficient to consider a Gaussian state (see discussion in \cite{noi})
peaking the (discrete variables representing the) boundary geometry (encoded in the spin
network labels $j_l$) around a given semiclassical configuration. Choosing the
equilateral configuration as the background geometry of the boundary 4-simplex, the data
encoded by $\Psi_0[j_l]$ are the value $2j_0$ of the area of the ten triangles, representing the intrinsic
curvature, and the value $\theta=\arccos(-\f14)$ of the ten dihedral angles, representing the extrinsic
curvature. In particular in \cite{RovelliProp}, the following ansatz was taken:
\equ\label{psi0}
\Psi_0[j_l]=\exp\left\{-\f{1}{2j_0}\sum_{ll'}\alpha_{ll'} \, \d j_l \, \d j_{l'} + i \theta \sum_l (2j_l+1) \right\},
\nequ
where $\d j_l = j_l-j_0$, and $\alpha_{ll'}$ is a $\mathbbm C$-valued (non diagonal) matrix
that can be fixed by comparing the leading order of \Ref{W2} with
(a suitable discretisation of) the conventional free propagator \cite{RovelliProp}.
Using \Ref{psi0}, we have $\f{\mean{\Delta j_l}}{\mean{j_l}} = \f1{\sqrt{j_0\alpha_{ll}}}$
and
$\f{\mean{\Delta \phi_l}}{\mean{\phi_l}}= \f1\theta\sqrt\f{\alpha_{ll}}{{j_0}}$, so that
\Ref{semi} is satisfied in the $j_0\mapsto\infty$ limit.

It was assumed that the phase of \Ref{psi0} suppresses the term $D(j_l)$.
Then, using \Ref{asymp} and \Ref{psi0}, in \cite{RovelliProp} it was showed that the leading order of \Ref{W2bis}, in the large spin limit, is
\equ\label{WLO}
W\sim\f1{j_0}.
\nequ
This shows that the leading order of the components of the 2-point function behave as
expected: rescaling the boundary geometry where the two points lie, $W_{ab}$ scales as
$1/|x-y|^2$.

However this evaluation relies on a
number of assumptions. In particular the use of \Ref{asymp}, and
the suppression of the degenerate configurations term $D(j_l)$.
It would thus be important to confirm the analytic result
\Ref{WLO} with numerical simulations of \Ref{W2bis}, but this has not been achieved so far
\cite{Dan}, due to the high complexity of the sum.
In the rest of this paper we address this issue, and we prove \Ref{WLO} in a way that has been verified
numerically \cite{Dan}. In particular, we do not use \Ref{asymp},
and we prove that the degenerate configurations plaguing the BC model are indeed suppressed.
First of all, notice that using the integral representation \Ref{cicciopanza} of the spinfoam amplitude, and swapping the integrals with the sums, we can rewrite \Ref{W2bis} as
\equ\label{W3}
W_{ab} = \f1{j_0^4}\f1{\cal N} \int \prod_n dg_n \sum_{j_l}
\ \prod_l  {\chi}_{j_l}(g_{s(l)}^{-1}g_{t(l)})
\ \Psi_0[j_l] \ {\mathbbm h}(j_a) \ {\mathbbm h}(j_b).
\nequ

The key idea is to use a boundary state $\Psi_0[j_l]$, so that we can perform
\emph{exactly} the sums in \Ref{W4}. To do so, recall that the kernel is nothing but the
$\SU(2)$ character. Then, to be able to perform the sum, it is sufficient to
have a state $\Psi_0[j_l]$ with a well--defined Fourier transform.
This is what we do in the next Section.

\section{The new boundary state}
\label{SectionBoundary}

As we showed above the relevant group for 4d Riemannian GR without matter is simply $\SU(2)$;
indeed, \Ref{W2} uses only $\SU(2)$ spins. This has interesting consequences,
as $\SU(2)$ is also the relevant group for 3d Riemannian GR. Therefore
we can apply to the Barrett--Crane model in 4d the same techniques
developed to study the graviton propagator in the Ponzano--Regge model
for 3d quantum gravity in \cite{noi2}. In particular, we consider the following new ansatz for the boundary state,
\eqa\label{psi}
\Psi_0(j_l)&=&\prod_l \psi_0(j_l), \\
\label{psicool}
\psi_0(j)&=&\f{ I_{|j-j_0|}(\f{j_0}{\alpha})
-I_{j+j_0+1}(\f{j_0}{\alpha})}{\sqrt{I_{0}(\f{2j_0}{\alpha})
-I_{2j_0+1}(\f{2j_0}{\alpha})}}
\,\cos(d_j \theta).
\neqa
Here the $I_n(z)$ are the modified Bessel functions of the first kind, and
$\alpha\in \mathbbm R^+$ is a free parameter.\footnote{With respect to \cite{noi2},
there is an important difference in the phase, which is given by $d_j \theta$ here and by $d_j \theta/2$
in the 3d case. The reason for this lies in the asymptotics \Ref{asymp}, which reproduce the
Regge action with areas $A=d_j$, whereas the asymptotics for the 3d spinfoam amplitude reproduce
the Regge action with lengths $\ell=d_j/2$.}

Notice that this new boundary state factorises in link contributions and is real, differently
from \Ref{psi0}.
As shown in \cite{noi2}, the $j_0\mapsto \infty$ limit of \Ref{psicool} behaves as a Gaussian peaked
around $j_0$,
\equ\label{limit}
\psi_0(j)\simeq \sqrt[4]{\f\alpha{{j_0}\pi}}
\exp\{-\f\alpha{2{j_0}}(j-j_0)^2\}\cos(d_j\theta),
\nequ
and thus \Ref{psi} satisfies the semiclassical requirements \Ref{semi}. Thanks to the
asymptotic behavior \Ref{limit}, a boundary state constructed from \Ref{psicool} can be
used to reproduce the leading order \Ref{WLO} using the same procedure outlined in the previous Section.

However, the exact form \Ref{psicool} opens the way to a
new treatment, which allows us give a check of the procedure of \cite{RovelliProp}.
The key property of \Ref{psicool} which we need in the following is its Fourier transform (see
\cite{noi2}),
\equ\label{psitilde}
\widetilde\psi_0(g)=\sum_j \psi_0(j)\, \chi_j(g)= \sum_{\eta=\pm}
\f{e^{-\f{2j_0}{\alpha}\sin^2(\phi-\eta\theta)}}{2N\sin\phi}\sin \Big(d_{j_0}(\phi-\eta\theta)\Big).
\nequ
where
\equ
N = e^{-\f{j_0}{\alpha}} \sqrt{ I_{0}(\f{2j_0}{\alpha}) - I_{2j_0+1}(\f{2j_0}{\alpha})}
\nequ
is the normalisation with respect to the Haar measure. From the Gaussian shape of \Ref{psitilde},
it is clear that this new boundary state peaks the $\SU(2)$ class angle $\phi$ around
the values $\eta\theta$ modulus $\pi$. For later purpose, we introduce a sign variable $\sigma=\pm$
and write the peaks as
\equ
\overline\phi=\eta \theta +\f{1-\sigma}2\pi.
\nequ
As we show below, the different values of the signs $\eta$ and $\sigma$
give the same contribution to the 2-point function,
thus summing over them simply adds a trivial redundancy.

The fact that \Ref{psicool} admits a simple Fourier transform is at the heart of our construction
of the 2-point function as an integral over $\SU(2)$.
In particular, in the following we also need the convolution product between
$\psi_0$ and the field insertion $\mathbbm h$, $\psi_0 \circ \mathbbm h$. This can be easily evaluated, using
\equ
\n^2=\f1{\sin^2\phi}\, \p_\phi \, \sin^2\phi \, \p_\phi, \qquad
\n^2 \chi_j(\phi)=-4 \, j(j+1) \, \chi_j(\phi),
\nequ
to write
\eqa\label{conv}
{(\psi_0 \circ \mathbbm h)}(\phi) &=&
\sum_j {\chi}_j(\phi) \, \Psi_0(j) \, \left[ C^2(j) - C^2(j_0)\right] =
 \left[ -\f14\n^2 - j_0(j_0+1)\right]  \widetilde\psi_0(\phi)= \\\nonumber  && \hspace{-2.5cm} =
\sum_{\eta=\pm}\f{j_0}\alpha
\f{e^{-\f{2j_0}\alpha\sin^2(\phi-\eta\theta)}}{2N\sin\phi}
\sin\Big(d_{j_0}({\phi-\eta\theta})\Big)\sin 2(\phi-\eta\theta)
\Bigg[\cot 2(\phi-\eta\theta)-
\f{j_0}{\alpha} \, {\sin 2(\phi-\eta\theta)}
+{d_{j_0}}\,{\cot \Big({d_{j_0}}(\phi-\eta\theta)\Big)}\Bigg].
\neqa
For later use, let us rewrite this expression as
\equ\label{conv1}
{(\psi_0 \circ \mathbbm h)}(\phi) = \sum_{\eps=\pm} \sum_{\eta=\pm} \f{j_0}\alpha
\f{e^{-\f{2j_0}\alpha\sin^2(\phi-\eta\theta)}}{2N\sin\phi}
\,\eps\, e^{i\eps d_{j_0}({\phi-\eta\theta})}\sin 2(\phi-\eta\theta)
\Big[\cot 2(\phi-\eta\theta)-\f{j_0}{\alpha} \, {\sin 2(\phi-\eta\theta)}+{d_{j_0}}\,i \, \eps \Big].
\nequ

Before proceeding, let us add a few remarks on the new boundary state, which
make it particularly appealing.
\begin{itemize}
\item{The spin and the angle entering the state, which represent a discretised version of the conjugate
intrinsic and extrinsic curvature variables, are conjugate variables in a precise
mathematical sense: they are conjugate with respect to the $\SU(2)$ harmonic analysis.}
\item{The state is a real quantity. This is due to the fact that the
phase is given by a cosine, and not a single exponent, as in \Ref{psi0}. The fact that using
the cosine and not just a single exponent does not spoil the leading order of \Ref{W2} was
proved in \cite{noi2}. Furthermore, one could also consider a sine term, or equivalently
an $\SU(2)$ character. In 3d, this can be related to particle insertions \cite{noi2, noiciccia}.}
\item{The boundary state has an interacting structure between the variables
and the background; it reduces to a Gaussian only in the limit $j_0\mapsto\infty$. For this
reason it will also contribute to the corrections
to the free propagator. In particular, the 3d analysis of this state shows that
its contribution  interestingly reduces the magnitude of the next to leading order.}
\item{The boundary state \Ref{psi} is factorised into contributions from
single links of the boundary spin network. This makes its analysis much simpler than
\Ref{psi0}, but also means that it is straightforward to extend it to arbitrary triangulations.}
\item{In the definition \Ref{psi} we took the same $\alpha$ for all links. Nothing prevents us
from considering arbitrary configurations with a different parameter for each link. Because this parameter is
in a sense related to the knowledge of the boundary geometry, taking the same value for all links amounts to
a homogeneous description of the boundary state.}
\end{itemize}

\section{The 2-point function as an integral over $\SU(2)$}
\label{SectionProp}

Thanks to the factorisation of the new boundary state, we can
rewrite \Ref{W3} as
\equ\label{W4}
W_{ab} =
\f1{j_0^4}\f1{\cal N} \int \prod_n dg_n \ \prod_l {\cal I}_l(g_l),
\nequ
where $g_l = g_{s(l)}^{-1}g_{t(l)}$, and
\eqa\label{I}
{\cal I}_l(g)= \left\{\begin{array}{l}
\sum_{j_l} \, {\chi}_{j_l}(g)\,\psi_0[j_l] \equiv \widetilde\psi_0(g),
\qquad {\rm if}\ \ \ l\neq a,b, \\ \nonumber \\
\sum_{j_l} \, {\chi}_{j_l}(g_l)\,\psi_0[j_l] \, {\mathbbm h}(j_l) \equiv
{(\psi_0 \circ \mathbbm h)}(g)
\qquad {\rm if}\ \ \ l=a,b, \end{array}\right.
\neqa
are given respectively by \Ref{psitilde} and \Ref{conv}.
This expression gives the graviton propagator as an integral over $\SU(2)$.

The choice of the adge amplitude $A_e(j_f)=(\prod_f d_{j_f})^{-1}$ made above
simplifies the computation of the two Fourier transforms.
More generically, if $A_e(j_f)$ does not couple the spins
and is simply a power of $\prod_f d_{j_f}$, $W_{ab}$ still factorises as
above, with the extra powers of $d_{j_l}$ acting as differential operators $\p_\theta$
in computing the Fourier transform. On the other hand, if the
edge amplitude couples the spins and introduces some $d_{j_1j_2j_3j_4}$
factors (see footnote \ref{d4}), the situation is different.
For a single 4-simplex, these factors can always be compensated
by introducing suitable counter--factors in the boundary state, and thus
the form \Ref{W4} can be restored. For multi--simplices configurations, this
will not work. However, for arbitrary triangulations it is not straightforward to obtain the integral representation
\Ref{W4} of the graviton, even with the simplest choice $A_e(j_f)=(\prod_f d_{j_f})^{-1}$.
In fact, the one--to--one correspondence between $\SU(2)$ characters and
factors \Ref{psicool} of the boundary state is in general lost. This makes it harder to
perform the sums explicitly. We postpone the study of arbitrary triangulations to future work,
nonetheless let us notice here that the correspondence is preserved by any
$n$-valent vertex, as long as the vertex amplitude is provided by the relativistic spin network \Ref{eps}.
Therefore, an integral expression like \Ref{W4} can be
obtained for any triangulation that can be coarse grained to a single $n$-valent vertex.

To explicitly evaluate the perturbative expansion in $j_0$, it is convenient to use the
measure
\Ref{ten} in terms of the ten angles. Notice that the factors $\sin{\phi_l}$ in \Ref{ten} simplify with the ones
coming from $\tilde\psi_0(g)$ and $(\psi_0 \circ \mathbbm h)(g)$, and we can write \Ref{W4} as
\equ\label{W5}
W_{ab} =
\f1{j_0^4}\f1{\cal N}\sum_{\eps_l=\pm}\sum_{\eta_l=\pm} \int \prod_l d\phi_l \ \d\Big(\det G_{IJ}\Big)
\prod_l \eps_l\,e^{i\eps_l\,d_{j_0}(\phi_l-\eta_l\theta)} \
{\mathbbm h}(\phi_a) \ {\mathbbm h}(\phi_b) \  e^{-\f{2j_0}\alpha \sum_l \sin^2(\phi_l-\eta_l\theta)},
\nequ
where we have absorbed the constant $(2N)^{10}$ in the normalisation, and
we have introduced the notation
\equ\label{ins2}
{\mathbbm h}(\phi_a) = \f{j_0}\alpha\,
\sin2(\phi_a-\eta_a\theta)\Big[ \cot2(\phi_a-\eta_a\theta)-\f{j_0}\alpha\,\sin2(\phi_a-\eta_a\theta)
+d_{j_0}\, i\, \eps_a \Big].
\nequ
The expression \Ref{W5} is the contribution to the graviton propagator coming from a
single 4-simplex.

Let us stress that \Ref{W5} is an integral over $\SU(2)$ with no
sums involved, as opposed to \Ref{W2bis}. This result is particularly important from a numerical point of view. The
integral is in fact much easier to handle numerically than the sums, and the
formulation provided here gives substantial progress in the numerical simulations \cite{Dan}. As
far as the sum over the $\eta$ signs are concerned, we show below that each configuration
gives the same contribution, so that the sum gives a trivial redundancy that can be
reabsorbed in the normalisation.

To define the perturbative expansion in $\lp$ we can use the parameter $j_0$
as above. Notice that $j_0$ enters the exponential of the integrand.
Therefore the leading order in the $j_0\mapsto \infty$ limit can be computed using
the saddle point approximation of the integral. We do so in the next Section.

\section{The saddle point approximation}
\label{SectionSaddle}

The study of the leading order and corrections to the 2-point function formulated as a
group integral requires the analysis the saddle point (or stationary phase) approximation
of the following type of integral:
$$
\int d\mu[\phi_l]\;
F(\phi_l)\;\sum_{\eta_l=\pm} e^{i\sum_l \eps_l d_{j_0}(\phi_l-\eta_l\theta)}\;
e^{-\f{2j_0}{\alpha}\sum_l\sin^2(\phi_l-\eta_l\theta)},
$$
where $\eps_l=\pm$ are signs, $F(\phi_l)$ an arbitrary function (with no dependence on
$j_0$) and the measure $d\mu(\phi_l)$ is defined in \Ref{ten}.
For large $j_0$ that integral is dominated  at the fixed points of the
action
\equ\label{action}
S[\phi_l, \kappa]=\sum_l \left[\f{2}{\alpha}\sin^2(\phi_l-\eta_l\theta) - 2i\eps_l(\phi_l-\eta_l\theta)\right]
-i\f\kappa{j_0}\, \det G_{IJ},
\nequ
where we used a Lagrange multiplier $\kappa$ to introduce the constraint $\det G_{IJ}=0$.
To compute the equations of motion, notice that
$
\f{\p \det G_{IJ}}{\p \phi_l}
\,=\, -2 \, \sin\phi_l \, \Lambda^{(l)},
$
where $\Lambda^{(l)}$ is the determinant of the off--diagonal $l$-th minor of $G_{IJ}$, obtained removing the
line and column corresponding to one of the two $\cos\phi_l$ appearing in it.

Then the equations of motion read:
\eqa\label{motion} \left\{ \begin{array}{l}
\f{2}{\alpha}\sin2(\phi_l-\eta_l\theta)-2i\eps_l
\,=\,
-2i\f\kappa{j_0} \, \sin\phi_l \, \Lambda^{(l)} \qquad \forall l,
\\ \\ \det G_{IJ}=0. \end{array} \right.
\neqa
The only real solution to this set of equations is:
\be\label{saddle}
\overline{\phi_l} =\eta_l \theta + \f{1-\sigma_l}2 \pi,
\qquad \overline{\eps_l}=\eps \, \eta_l
\ee
where $\eps=\pm$ is a global sign (independent of the link $l$), and it requires the symmetry property
$\sigma_l = \sigma_{IJ} = \sigma_I \sigma_J$, with $\sigma_I$ and $\sigma_J$
independent signs. We have $\sin\overline{\phi_l}= \eta_l \sigma_l \sin\theta$ and
$\Lambda^{(l)}(\overline{\phi_l}) = \sigma_l \, \Lambda_0$,
$\Lambda_0:=\cos\theta \, (1-\cos\theta )^3$. Consequently, the Lagrange multiplier takes the value
\equ\label{kappa}
\overline\kappa = \eps \, \f{j_0}{ \sin\theta \, \Lambda_0}.
\nequ
The action evaluated on the fixed
point \Ref{saddle} is purely real and vanishes, $S[\overline{\phi_l},\overline\kappa]=0$. If
$\eps_l\neq \overline{\eps_l}$,
there is no real solution. The saddle point approximation selects two
specific terms out of the sum over all possible sign assignments.

More generally, the first part of the action, $S_0[\phi, \kappa]=-\sum_l 2i \eps_l\phi_l
-i \f\kappa{j_0} \, \det G_{IJ}$, is the standard action governing the asymptotics of the (equilateral)
$\{10j\}$ symbol. It has two non--degenerate fixed points $\phi_l=\theta$ and
$\phi_l=-\theta$, but also degenerate fixed points at $\phi_l=0,\pi$.
These degenerate fixed points dominate the asymptotic behavior of the $\{10j\}$ symbol
\cite{asymptjohn, asymptlaurent,asymptbaez}, and are responsible for the $D$ term in \Ref{asymp}.
However, we see that their contribution to the 2-point function is suppressed: in
\Ref{W5}, there is an additional Gaussian weight, which is maximized by $\phi_l=\theta$
but not by $\phi=0, \pi$. Therefore it kills the degenerate configurations and peaks the
asymptotics around the non--degenerate semiclassical configuration. This
provides the mechanism to neglect the $D$ term, as done in \cite{RovelliProp}.

We have not shown that there is no complex solution to these equations although we believe there is
none.\footnote{At least, it is fairly straightforward to check that there is no purely imaginary
solutions $\phi_l\in i\R$.} In particular, there could be fixed 
points close or along the imaginary line $\phi_l-\eta_l\theta\in i\R$, for which the
Gaussian would blow up. Such points can be easily found if $\det G_{IJ}\neq 0$ is allowed,
but otherwise we have not been able to find any. The results of our partial analysis
suggest that the constraint $\det G_{IJ}=0$ protects the integral from the presence of
exponentially enhanced fixed points, a mechanism that could be relevant also for the 
asymptotics of the \1 symbol alone.

Before proceeding, let us add an important remark.\footnote{We thank our referee for pointing this out.}
After the identification of the fixed points, one usually computes the integral along the
complex  contour such that the phase (namely the imaginary part of $S$)
varies the least. This steepest descent method is implemented imposing that the imaginary part of the second
derivative of $S$ vanishes. However, in the following we will use the 
real contour, which does not satisfy this requirement
(as can be immediately seen from the explicit expression of the second derivatives in \Ref{motion}).
Nevertheless, as we mention at the end of this section, a numerical check has been performed independently from our
analysis and the fit is (at least) of two decimals. This means that most likely there are no other
(relevant) fixed points and that the real contour is enough for the study of the correlations.

The fact that complex fixed points, if they do exist, are not relevant for our
analysis may be due to the following reasons. First, 
notice that we are summing over all signs $\eta_l$, in such a way that only the real part of the integral matters
(see below), a fact that might lessen the relevance of the phase. Then, recall that 
the precise normalisation of the integral is not relevant
but only its ratio with $\cal N$, the integral without the ${\mathbbm h}$ insertions.
Finally, we are merely computing the first order of the large $j_0$ asymptotics of $W_{ab}$. 
The existence of complex fixed points might become relevant when computing
the higher order corrections. We leave this question open for future investigations.

Based on this preliminary analysis of the stationary points, we are now ready to state
the main result of this paper on the asymptotics of the 4-simplex 2-point function.

\begin{theo}
The leading order of the large $j_0$ expansion of \Ref{W5} is
\equ\label{theo}
W_{ab}(j_0) = \f{f_{ab}(\alpha, \theta)}{j_0},
\nequ
where $f_{ab}$ has only three independent entries.
\end{theo}
\noindent These correspond to the
three independent projections of the graviton propagator along
couples of normals belonging to the same triangles (see discussion in Section \ref{Section2}), and should not
be confused with the physical independent components.
The explicit structure of $f_{ab}(\alpha)$, reported in the proof below, is rather complicate,
reflecting the complicate geometrical structure of a 4-simplex, even in the equilateral case.

\begin{proof}
The proof goes as follows.
\begin{enumerate}
\item{We rewrite \Ref{W5} introducing a Lagrange multiplier
$\kappa$ for the constraint $\det G_{IJ}$,
\equ\label{W6}
W_{ab} =
\f1{j_0^4}\f1{\cal N}\sum_{\eta_l=\pm}\sum_{\eps_l=\pm} \int d\kappa \int \prod_l d\phi_l \
\left(\prod_l \eps_l \right) \ e^{-j_0 S[\phi_l, \kappa] +i \sum_l\eps_l (\phi_l-\eta_l\theta)}
\ {\mathbbm h}(\phi_a) \ {\mathbbm h}(\phi_b),
\nequ
where $S[\phi_l, \kappa]$ is given by \Ref{action}.
The normalisation $\cal N$ is given by the same
quantity above without the insertions $\mathbbm h(\phi_a) \ {\mathbbm h}(\phi_b)$ and
without the constant factor $j_0^{-4}$.
In the calculations below, we will use $\cal N$ to reabsorb a number of overall constants,
without affecting the final result.

The expansion parameter $j_0$ enters the exponent in front of
the action $S$, and the field insertions $\mathbbm h$.
The saddle point of the action is
\equ\label{points}
\phi_l=\overline{\phi_l}, \qquad \eps_l=\overline{\eps_l}, \qquad \kappa=\overline\kappa,
\nequ
where $\overline{\phi_l}$ and $\overline{\eps_l}$ are given in \Ref{saddle}, and $\overline{\kappa}$ in \Ref{kappa}.
}
\item{
To compute the leading order of \Ref{W6}, we expand the action to second order
around the saddle point (where both the zeroth and the linear orders vanish), and evaluate the rest at lowest order.
In particular, at the saddle point we have
\equ\label{sign1}
\prod_l \eps_l{\Big|_{\rm saddle}} =\prod_l \eps \, \eta_l \equiv \prod_l \eta_l,
\qquad
e^{i \sum_l\eps_l (\phi_l-\eta_l\theta)}{\Big|_{\rm saddle}}=
\prod_l e^{i\eps\,\eta_l\f{1-\sigma_l}2 \pi} = \prod_l \sigma_l.
\nequ
Consequently, we obtain the global sign
$\prod_l \eta_l\,\sigma_l$. When we perform the sum over the $\eta_l$ signs in \Ref{W6}, we obtain
identically zero unless
we have the matching $\eta_l=\sigma \,\sigma_l$ $\forall l$, where $\sigma$ is an irrelevant
overall sign.
In the following, we take $\sigma_l = \eta_l$.

The leading order of \Ref{W6} is thus
\eqa\label{W7}
W_{ab} \simeq \f1{j_0^4}\f1{\cal N}\sum_{\eta_l=\pm}\sum_{\eps=\pm} \int d\d\kappa \int \prod_l d\d\phi_l \
e^{-\f{j_0}2 \Big(\f{\p^2 S}{\p\phi_l \p\phi_{l'}}\d\phi_l \d\phi_{l'}
+ \f{\p^2 S}{\p\phi_l \p\kappa}\d\phi_l \d\kappa  \Big)}
\ {\mathbbm h}(\overline{\phi_a}+\d \phi_a) \ {\mathbbm h}(\overline{\phi_b}+\d \phi_b).
\neqa
}
\item{In principle, the value of the field insertions at the saddle point is enough
for the leading order. However, we have $\mathbbm h(\overline\phi)\equiv 0$, thus we need
to expand the field insertions. This can be done in two steps. First, we keep only
the terms in $j_0^2$ in \Ref{ins2}, thus
\equ
\mathbbm h(\phi_a) \simeq
(\f{j_0}\alpha)^2 \,\sin2(\phi_a-\eta_a\theta) \Big[2\alpha\,i\,\eps_a
-\sin2(\phi_a-\eta_a\theta) \Big].
\nequ
Then, we expand $\phi=\overline\phi+\d \phi$. We have
$\sin\Big((1-\sigma_a)\pi+2\d\phi_a\Big)= \sin 2\d\phi_a\simeq 2 \, \d\phi_a$.
Using also $\eps_a=\eps \, \eta_a$, we can write
\equ\label{hexp2}
\mathbbm h(\overline\phi_a+\d \phi_a)\simeq  (\f{j_0}\alpha)^2 \,2\, \d\phi_a\,
\Big[2\alpha\,i\,\eps \, \eta_a -2 \,\d\phi_a \Big].
\nequ
However, these two terms are of different orders. In fact,
notice that \Ref{W7} is a Gaussian integral with width proportional to $1/\sqrt{j_0}$. Therefore
$\d\phi^2\sim 1/j_0$, so the second term in \Ref{hexp2} can be neglected, and finally
\equ
\mathbbm h(\overline\phi_a+\d \phi_a)\simeq  4\alpha\,i\,(\f{j_0}\alpha)^2 \,\eps \, \eta_a\, \d\phi_a.
\nequ
}
\item{The second derivatives of the action have the following form,
\equ
\f{\p^2 S}{\p\phi_l \p\phi_{l'}}\Big|_{\rm saddle} = \eta_l \, \eta_{l'} \, A^\eps_{ll'},
\qquad
\f{\p^2 S}{\p\phi_l \p\kappa}\Big|_{\rm saddle} = \f{2i}{j_0}\, \eta_l \, \sin\theta \, \Lambda_0.
\nequ
The explicit form of $A^\eps_{ll'}$ is given in the Appendix, and it
satisfies $A^-_{ll'}= \overline{A^+_{ll'}}$.

Notice that the constant factor $\sin\theta\, \Lambda_0$ can be neglected rescaling the
definition of $\d\kappa$, and then reabsorbing it in the normalisation ${\cal N}$ of \Ref{W7}.
}
\item{The leading order of \Ref{W6} is thus the following Gaussian integral,
\eqa\label{W8}
W_{ab} &=& -\f1{\cal N}
\f{16}{\alpha^2}\sum_{\eta_l=\pm}\sum_{\eps=\pm}
\eta_a \, \eta_b
\int d\d\kappa \int \prod_l d\d\phi_l \ \d\phi_a \, \d\phi_b \,
\exp\left[- \eta_l \eta_{l'} \f{j_0}2 A^\eps_{ll'} \d\phi_l \d\phi_{l'}
-i\eta_l \d\phi_l \d\kappa \right] = \no \no
&=& -\f{16}{\alpha^2}\f1{Z(0)}\,\f{\p^2}{\p J_a \p J_b}\, Z(J)\Big|_{J=0},
\neqa
where we have introduced the generating functional
\equ\label{gen}
Z(J) = \sum_{\eta_l=\pm}\sum_{\eps=\pm} \int d\d\kappa \int \prod_l d\d\phi_l \
\exp\left[- \eta_l \eta_{l'} \f{j_0}2 A^\eps_{ll'} \d\phi_l \d\phi_{l'}
-i\eta_l \d\phi_l \Big(\d\kappa + i J_l \Big)\right].
\nequ
}
\item{The generating functional is a Gaussian integral that can be evaluated straightforwardly.
To maintain the explicit symmetry of the problem, we perform first the integral over the ten angles, obtaining
\eqa
Z(J) = \sum_{\eta_l=\pm}\sum_{\eps=\pm}\f{(2\pi)^{5}}{j_0^5 \, \sqrt{\det A^\eps_{ll'}}}
\int d\d\kappa \ \exp\left[-\f1{2j_0} \sum_{ll'}(\d\kappa + iJ_l) (A_\eps{}^{-1})_{ll'} (\d\kappa + iJ_{l'})\right].
\neqa
Here we used the fact that $\det \Big( \eta_l \, \eta_{l'} \, A^\eps_{ll'} \Big)\equiv \det A^\eps_{ll'}$.
Observe that consequently the sums over the $\eta_l$ signs give a trivial redundancy
$2^{10}$.
The remaining integral is also straightforward, and we finally obtain
\equ
Z(J) = 2^{10} \sum_{\eps=\pm} \f{(2\pi)^{5}\,\sqrt{2\pi j_0}}
{j_0^5  \, \sqrt{\det A^\eps_{ll'} \, \sum_{ll'}(A_\eps{}^{-1})_{ll'}}} \
\exp\left[\f1{2j_0} \left(\sum_{ll'}(A_\eps{}^{-1})_{ll'} J_l J_{l'}
-\f{\Big(\sum_{ll'}(A_\eps{}^{-1})_{ll'}
J_l\Big)^2}{\sum_{ll'}(A_\eps{}^{-1})_{ll'}} \right)\right].
\nequ

Because $A^-_{ll'}= \overline{A^+_{ll'}}$, the sum over the $\eps$ sign amounts to taking (twice) the real part
of the summand.
Reabsorbing the irrelevant constants in the normalisation $Z(0)$ and defining $A_{ll'} = A_{ll'}^+$, we can then write
\equ
Z(J) = {\rm Re} \, \left\{\f{1}
{{\sqrt{\det A_{ll'} \, \sum_{ll'}(A^{-1})_{ll'}}}} \
\exp\left[\f1{2j_0} \left[\sum_{ll'}(A^{-1})_{ll'} J_l J_{l'} -
\f{\Big(\sum_{ll'}(A^{-1})_{ll'} J_l\Big)^2}{\sum_{ll'}(A^{-1})_{ll'}} \right)\right]\right\}.
\nequ
}
\item{As shown in the Appendix,
the matrix $(A^{-1})_{ll'}$ satisfies $\sum_{l'}(A^{-1})_{ll'} = f_1(\alpha)$ $\forall l$,
where $f_1(\alpha) = \f{\sqrt{15}}{2(\f{2\sqrt{15}}\alpha + 5i)}$,
and thus $\sum_{ll'}(A^{-1})_{ll'} = 10 f_1(\alpha)$.

Using this, we can finally write
\equ\label{W9}
W_{ab} = -\f1{\cal N}\f{16}{\alpha^2 \, j_0} \,{\rm Re} \, \left\{
\f{A_{ab}^{-1}-\f1{10}f_1(\alpha)}{\sqrt{\det A \ 10 f_1(\alpha)}}\right\},
\qquad
{\cal N} = {\rm Re} \, \left\{ \f1{\sqrt{\det A \  10 f_1(\alpha)}}\right\}.
\nequ
}
\item{
The theorem is proved with
\equ
f_{ab}(\alpha) = -\f{16}{\alpha^2}\f{{\rm Re} \, \left\{
\f{A_{ab}^{-1}-\f1{10}f_1(\alpha)}{\sqrt{\det A \ f_1(\alpha)}}\right\} }
{{\rm Re} \, \left\{ \f1{\sqrt{\det A \ f_1(\alpha)}}\right\} }
\nequ
}

\end{enumerate}
\end{proof}

If seen as a matrix, $f_{ab}(\alpha)$ has only three independent components, as it is
shown in the Appendix. Geometrically, these are related to the three cases (i) when $a$
and $b$ are the same triangle, (ii) when they share a side (they are ``adjacent'' in the
4-simplex), or (iii) when they share only one point (they are ``opposite'' in the
4-simplex).

For instance, choosing the value $\alpha=1/2$, we have
\equ
W_{\rm adj}(j_0) \simeq \f{1.21}{j_0}, \qquad W_{\rm opp}(j_0) \simeq -\f{0.61}{j_0}.
\nequ
Remarkably, these values can be confirmed by a numerical analysis of \Ref{W6} \cite{Dan}.

\section{Conclusions and Outlook}
We have introduced a new boundary state to construct the graviton propagator. This new
boundary state is given in \Ref{psi}, and in the large spin limit it reduces to the same
Gaussian considered in \cite{RovelliProp}. Using the new state we were able to write the
(contribution from a single
4-simplex to the) graviton propagator as an integral over $\SU(2)$. This is explicitly given in \Ref{W5}.
The result is very useful from the point of view of numerical analysis (see \cite{Dan}), which is
significantly simplified with respect to the expression of \cite{RovelliProp}.
Also, the integral expression of the graviton
allows to recast the $\lp$ expansion into the saddle point expansion of the integral.
Here we evaluated the leading order, given in \Ref{theo}.
The reason why the degenerate configurations of the Barrett--Crane model can be neglected emerges
clearly: they correspond to non absolute minima of the action. As such, they do not enter
the perturbative expansion.

These results provide a starting point for further developments. Let us mention a few which we regard as
particularly interesting:

\begin{itemize}

\item Compute the next to leading order correction. This is a formidable task, but the integral expression here
obtained allows a precise setting to do it.
One has to consider higher orders in the expansion of the action \Ref{action} around the saddle point, as
well as higher orders in the expansion of the field insertions. Furthermore, we expect different choices
of edge amplitudes to affect the next to leading order (as it happens in 3d \cite{noi2}), and
thus computing it should allow to discriminate between the different versions of the Barrett--Crane model: different
choices of coupling between the 4-simplices will lead to different higher orders.
From this point of view, the fact that the expression can also be studied numerically
provides a crucial support.

\item Compute the angle correlations. This permits to reconstruct the whole tensorial
structure of \Ref{W}, and check the number of physical degrees of freedom.
Up to now, we have looked at correlations
between the spins $j$ associated to two triangles of the 4-simplex. These area correlations correspond to
components of the type $W_{aabb}(x,y)$. To access the remaining components, we need to
look at observables involving four different triangles of the 4-simplex. This
corresponds to correlations between dihedral angles, \emph{i.e.} between two intertwiners within the
4-simplex. Looking at such correlations was also suggested in \cite{gft} in order to
study the possibility of long--range correlations in BC--like spinfoam models.

\item The boundary state has a parameter, $\alpha$, in some sense related to the knowledge of the boundary geometry.
Can we consider also the angle $\theta$ as an external  parameter, to
be related to the choice of triangulation? This is an interesting question that can be addressed
looking at what happens if we choose an angle $\tl{\theta}\ne
\theta=\arccos (-1/4)$ in the boundary state \Ref{psi}. The phase of the boundary state would not match
anymore the oscillating term of the \1 symbol and the saddle point approximation would
fail. This means that the single 4-simplex would not give the leading order of the
correlation anymore: a different spacetime triangulation is needed to obtain the $1/j_0$ behaviour.
For example, for an angle $\tl{\theta}= n\theta$ we can imagine
a configuration with $n$ 4-simplices to dominate over the
single 4-simplex configuration, and possibly give the $1/j_0$ behaviour.
Notice that the new triangulation giving the leading order would depend on the group field theory
coupling constant $\lambda$ and the choice of coupling between 4-simplices (edge amplitude).
More work on many 4-simplices configurations, following the
original analysis of \cite{noi}, is needed in order to understand these situations and
possibly highlight the precise physical role of $\lambda$.

\item Consider the Lorentzian case. Now the relevant group is the
non-compact Lorentz group $\SO(3,1)\sim\SL(2,\C)$. We expect the procedure proposed in this paper
to extend directly, once the harmonic analysis for the Lorentz group is used.
In particular, the Gaussian $\exp(-\sin^2\phi)$ for $\SU(2)$ used in the boundary state
should be replaced by the analogous
$\exp(-\sinh^2\phi)$ for $\SL(2,\C)$, or maybe by the simpler Gaussian $\exp(-\phi^2)$.
The latter case could further simplify the analysis of the group integrals.

\end{itemize}

\section*{Acknowledgments}

We thank John Baez, Dan Christensen and Carlo Rovelli for useful discussions and encouragement.
Research at Perimeter Institute is supported in part by the Government
of Canada through NSERC and by the Province of Ontario through MEDT.

\appendix

\section*{Appendix: Geometry of a 4-simplex}

Where more convenient, we use the double index notation for the links:
$l\equiv IJ$, where $I$ and $J$ are the two nodes linked by $l$.

Let us consider five 4d unit vectors $\hat{N}_I \in {\cal S}^3$, $I=1\ldots 5$, and introduce
the ten angles defined by their scalar products, $\cos\phi_{IJ}=\hat{N}_I\cdot\hat{N}_J$,
with the convention $\phi_{II}=0$. Finally, we define the $5\times 5$ Gram matrix,
$G_{IJ}=\cos\phi_{IJ}$.
These five vectors are not linearly independent, so we can find $v_I\,\in\R^5$ such
that:
\be
\sum_{I=1}^5 v_I \, \hat{N}_I=0.
\label{closure}
\ee
This means that the 5-vector $v_I$ is a null vector for the Gram matrix $G_{IJ}$. In
particular, we get a constraint on the angles $\phi_{IJ}$:
\be
\forall J,\quad \sum_I v_I \, \hat{N}_I\cdot\hat{N}_J\,=\,0
\quad \Rightarrow\quad
\det G_{IJ} =0.
\label{closure2}
\ee

This constraint can be interpreted geometrically as follows. The five unit vectors define a unique geometric
4-simplex (embedded in $\R^4$) up to a global scale (4-volume of the simplex). They are
the (outward) normals to the five tetrahedra of the 4-simplex. The closure condition of the
4-simplex reads exactly as \Ref{closure} with the $v_I$ being the (oriented)
3-volumes of the tetrahedra. Furthermore, we can differentiate the equation
\Ref{closure2} and contract it with the null vector. This gives:
\be\label{pluto}
\sum_{I,J} v_I \, v_J \,
\sin\phi_{IJ} \, {\rm d}\phi_{IJ}
\,=\,0.
\ee
Next, recall the well known relation
$3 \, v_I \, v_J \, \sin\phi_{IJ}
\,=\,4 \,{\cal V} \, A_{IJ},$
where ${\cal V}$ is the 4-volume of the simplex and $A_{IJ}$ the area of the
triangle shared by the two tetrahedra $I$ and $J$. This relation allows to write \Ref{pluto}
in a simple geometric form as
\be
\sum_{I\ne J} A_{IJ}\,{\rm d}\phi_{IJ}
\,=\,
0,
\ee
which is  the Schl\"afli identity.


On the other hand, directly differentiating the $\det G_{IJ}=0$ constraint gives:
\be\label{diffG}
\sum_{I<J} \sin\phi_{IJ} \, \Lambda^{(IJ)} \,{\rm d}\phi_{IJ}=0,
\ee
where $\Lambda^{(IJ)}$ is the off--diagonal minor obtained by removing the $I$th line and
$J$th column from the Gram matrix $G_{IJ}$. In particular, this means that the minors
$\Lambda^{(IJ)}$ are related to the areas $A_{IJ}$, up a global scale factor $\kappa$:
\be
\Lambda^{(IJ)}
\,=\, \kappa\,\f{A_{IJ}}{\sin\phi_{IJ}}.
\ee
This relation was used in Section II.

The minors $\Lambda^{(IJ)}\equiv\Lambda^{(l)}$ play a major role in evaluating the components of
the graviton propagator. In fact, their derivatives enter the second derivatives of the action
\Ref{action}. Using  $\sin\overline{\phi_l}= \eta_l \sigma_l \sin\theta$ and
$\Lambda^{(l)}(\overline{\phi_l}) = \sigma_l \, \Lambda_0$,
$\Lambda_0:=\cos\theta \, (1-\cos\theta )^3$,
at the saddle point \Ref{points} we have

\equ
\f{\p^2 S}{\p\phi_l\,\p\phi_{l'}}\Big|_{\rm saddle} =
\f4\alpha\cos2(\phi_l-\eta_l\theta)\,\d_{ll'}
+2i\f\kappa{j_0}\,\Big(\cos\phi_l\,\Lambda^{(l)}\,\d_{ll'}+ \sin\phi_l\, \f{\p\Lambda^{(l)}}{\p\phi_{l'}} \Big)
\Big|_{\rm saddle}
\equiv \eta_l\,\eta_{l'}\,A^\eps_{ll'},
\nequ
where
\equ\label{defA}
A^\eps_{ll'}:=\left(\f4\alpha+2i\eps\cot\theta\right)\, \d_{ll'} + 2i\eps \, \f1{\Lambda_0}\, \f{\p \Lambda^{(l)}}{\p \phi_{l'}}\Big|_{\phi=\theta},
\nequ
and
\equ
\f{\p^2 S}{\p\phi_l\,\p\kappa}\Big|_{\rm saddle} = 2i\f1{j_0}\sin\phi_l\,\Lambda^{(l)}\Big|_{\rm saddle}
=2i\f1{j_0}\eta_l\,\sin\theta\,\Lambda_0.
\nequ

Notice that $A_{ll'}^{-}\equiv \overline{A_{ll'}^+}$.

Evaluated at the saddle points, the derivatives of the minors have the following values,
\equ
\f{\p \Lambda^{(l)}}{\p \phi_{l'}}\Big|_{\rm saddle} =
\sigma_l \, \eta_l \, \f{\p \Lambda^{(l)}}{\p \phi_{l'}}\Big|_{\phi=\theta} =
\left\{ \begin{array}{ll}
-\sigma_l \, \eta_l \, \Lambda_0
\, \f{2\cos^2\theta + 3\cos\theta+1}{\sin\theta\cos\theta} \ \ &  {\rm if} \; {l=l'}, \\ \\
\sigma_l \, \eta_l \, \Lambda_0 \, \f{1+\cos\theta}{\sin\theta}
&  {\rm if} \; {l \;{\rm and}\; l' \;{\rm adjacient}}, \\ \\
0 & {\rm if} \; {l \;{\rm and}\; l' \;{\rm opposite}}.
\end{array}\right.
\nequ
Using these and the explicit value $\theta=\arccos(-\f14)$ of the angle, \Ref{defA} reads
\equ
A^\eps_{ll'}= \f6{\sqrt{15}}\Big( \f{2\sqrt{15}}{3\alpha} -\f{i}3 \eps \Big) \d_{ll'}
+\f{6}{\sqrt{15}}i\eps \, \left\{ \begin{array}{l}
2 \qquad  {\rm if} \; {l=l'}, \\ \\
1 \qquad  {\rm if} \; {l \;{\rm and}\; l' \;{\rm adjacient}}, \\ \\
0 \qquad {\rm if} \; {l \;{\rm and}\; l' \;{\rm opposite}}.
\end{array}\right.
\nequ
The matrix $A_{ll'}:=A_{ll'}^+$ has only three
independent entries, coming from the three independent components of the matrix $\f{\p \Lambda^{(l)}}{\p\phi_{l'}}$.
Consequently, each row of the matrix and of its inverse shows the same structure, which is summarised in the table
below. For convenience, we have introduced the quantity $\alpha' =\f{2\sqrt{15}}{3\alpha}$.

\begin{center}\framebox{
\begin{tabular}{c||c||c}
{\bf matrix $A_{ll'}$} & & {\bf inverse matrix $(A^{-1})_{ll'}$} \\ \hspace{4cm} & \hspace{4cm} & \hspace{5cm} \\
$\f{6}{\sqrt{15}}\left(\alpha'+i\f53\right)$  & along the diagonal &
$\f{\sqrt{15}}2\f{(3\alpha'+2i)(3\alpha'+11i)}{(3\alpha'+20i)(-7i+3\alpha')(3\alpha'+5i)}$ \\ & & \\
$\f{6}{\sqrt{15}}i$ & occuring six times & $-i\,\f{3\sqrt{15}}2 \f{(3\alpha'+11i)}{(3\alpha'+20i)(-7i+3\alpha')(3\alpha'+5i)}$ \\ & & \\
$0$ & occuring three times & $ i\, \f{3{\sqrt{15}}(3\alpha'+11i)}{(3\alpha'+20i)(-7i+3\alpha')(3\alpha'+5i)}$ \\
\end{tabular}}
\end{center}

Finally, we have
\equ
\det A_{ll'} = \f{1024}{759375}(3\alpha'+5i)(3\alpha'+20i)^4(-7i+3\alpha')^5,
\nequ
and
\equ
f_1(\alpha):=\sum_{l'}(A^{-1})_{ll'}
\equiv \f{\sqrt{15}}{2(\f{2\sqrt{15}}\alpha + 5i)} \qquad \forall l,
\nequ
from which $\sum_{ll'}(A^{-1})_{ll'} = 10 f_1(\alpha)$.


\end{document}